\documentclass[fleqn,usenatbib]{mnras}

\usepackage{newtxtext,newtxmath}

\usepackage[T1]{fontenc}

\DeclareRobustCommand{\VAN}[3]{#2}
\let\VANthebibliography\thebibliography
\def\thebibliography{\DeclareRobustCommand{\VAN}[3]{##3}\VANthebibliography}

\usepackage{graphicx}	
\usepackage{amsmath}	

\graphicspath{{./}{figures/}}

\title{Dynamics of Stellar Disk Tilting from Satellite Mergers}

\author[B. C. Dodge et al.]{
Benjamin C. Dodge,$^{1}$\thanks{E-mail: bendodge@alumni.princeton.edu}
Oren Slone,$^{1,2}$
Mariangela Lisanti$^{1,3}$
and Timothy Cohen$^{4,5,6}$
\\
$^{1}$Department of Physics, Princeton University, Princeton, NJ 08544, USA\\
$^{2}$Center for Cosmology and Particle Physics, Department of Physics, New York University, New York, NY 10003, USA\\
$^{3}$Center for Computational Astrophysics, Flatiron Institute, New York, NY 10010, USA\\
$^{4}$CERN, Theoretical Physics Department, Geneva, Switzerland\\
$^{5}$Theoretical Particle Physics Laboratory (LPTP), Institute of Physics, EPFL, Lausanne, Switzerland\\
$^{6}$Institute for Fundamental Science, Department of Physics, University of Oregon, Eugene, OR 97403, USA
}

\date{Accepted XXX. Received YYY; in original form ZZZ\hfill CERN-TH-2022-114}

\pubyear{2022}

\begin{document}
\label{firstpage}
\pagerange{\pageref{firstpage}--\pageref{lastpage}}
\maketitle

\begin{abstract}
The Milky Way's stellar disk can tilt in response to torques that result from infalling satellite galaxies and their associated tidal debris. 
In this work, we explore the dynamics of disk tilting by running $N$-body simulations of mergers in an isolated, isotropic Milky Way--like host galaxy, varying over satellite virial mass, initial position, and orbit. We develop and validate a first-principles understanding of the dynamics that govern 
how the host galaxy's stellar disk responds to the satellite's dark matter debris. We find that the degree of disk tilting can be large for cosmologically-motivated merger histories.  In particular, our results suggest that the Galactic disk may still be tilting in response to \emph{Gaia}-Sausage-Enceladus, one of the most significant recent mergers in the Milky Way's history.  These findings have implications for terrestrial direct detection experiments as disk tilting changes the relative location of the Sun with respect to dark matter substructure left behind by a merging galaxy.  
\end{abstract}




\section{Introduction} 
\label{sec:introduction}

Hierarchical structure formation dictates that galaxies like the Milky Way grow by merging with smaller satellite systems~\citep{1978MNRAS.183..341W}. These satellites are captured by the gravitational pull of the galaxy and experience extensive tidal mass loss as they orbit, until they are totally disrupted. The trail of debris left in their wake  builds up the accreted stellar halo~\citep{Johnston_1998,1999MNRAS.307..495H,2005ApJ...635..931B} and implies the existence of non-trivial phase-space substructure for the host galaxy's dark matter~(DM) halo~\citep{Diemand:2008in,Vogelsberger2011,Lisanti:2011as}.  The total mass of the DM that is left behind by any individual satellite can be comparable to, or even exceed, the mass of the host galaxy's stellar disk, leading to important dynamical effects.

The focus of this work is on one such effect, stellar disk tilting, which refers to changes in the direction of the disk's total angular momentum with time~\citep{1986MNRAS.218..743B,1989MNRAS.237..785O}. Unlike disk warping, tilting is most relevant for the inner part of the disk, which is tightly coupled due to self gravity and thus acts like a coherent body~\citep{2006MNRAS.370....2S}.  Using both isolated host-satellite simulations as well as cosmological hydrodynamic simulations, a variety of mechanisms have been identified that can cause disk tilting (e.g.~\cite{2017MNRAS.469.4095E}), including torques on the stellar disk from a satellite galaxy as it merges with the host~\citep{1975ApJ...202L.113O,1997ApJ...480..503H,1998ApJ...506..590S,1999MNRAS.304..254V,2004MNRAS.351.1215B,2006MNRAS.370....2S,2008MNRAS.389.1041R,villalobos2008simulations, 2009ApJ...700.1896K, 2012MNRAS.420.3324B,2022MNRAS.513.1867D}.  Other relevant mechanisms include DM halo triaxiality and figure rotation~\citep{1992ApJ...401..441D, 2004ApJ...616...27B,2007MNRAS.380..657B, 2012MNRAS.426..983D,2015MNRAS.452.2367Y, 2015MNRAS.452.4094D,Valluri_2021}, as well as gas accretion onto the stellar disk~\citep{2015MNRAS.452.4094D,2019MNRAS.488.5728E}.

Within the $\Lambda$~Cold Dark Matter~($\Lambda$CDM) paradigm, a galaxy like the Milky Way will experience a number of mergers as it evolves with time.  For example, at a redshift of $z<1$, an average of 5.1 (1.2) satellites with virial masses greater than $ 10^{10}$~$(10^{11})~M_\odot$ are expected to fall within the virial radius of a Milky Way--like host.\footnote{These estimates are based on merger trees generated using the \texttt{SatGen} code package~\citep{2021MNRAS.502..621J} as described in Sec.~\ref{sec:initial_conditions}, for hosts with virial masses at $z=0$ between $10^{12.1}~M_\odot$ and $10^{12.3}~M_\odot$.}  Understanding precisely how such a set of cosmologically-motivated  scenarios impact the degree of disk tilting for the Milky Way remains an open question. 

In this paper, we focus specifically on a Milky Way--like host and develop a first-principles understanding of the tilting dynamics between a satellite and the host's disk. We perform $N$-body simulations of mergers to study how the degree of disk tilting varies over a wide range of satellite initial conditions and find that the stellar disk typically precesses around and aligns with an axis that depends on these initial conditions. We also quantify the tilting rate as a function of the initial conditions and show that this rate may be significant and long-lasting for relevant merger histories.

The simulations run in this work track a satellite's orbital evolution in an isolated, isotropic halo.  Our results suggest that an individual satellite at infall redshift $z\sim1$ with virial mass larger than $\sim 10^{10}~M_\odot$ can lead to tilting rates greater than $\sim 0.3\,^\circ/\text{Gyr}$, depending on its orbital parameters.  This falls within the sensitivity range of the \emph{Gaia} satellite~\citep{2016A&A...595A...1G}, which will be able to determine changes to the orientation of the disk's angular momentum relative to an inertial reference frame defined by nearly a million observed quasars~\citep{Perryman:2014gka}. This suggests that more detailed studies are warranted.  For example, since our simulations are not cosmological, they do not include potentially important effects such as halo triaxiality and redshift-dependent halo and disk mass growth. Moreover, we only evolve one satellite at a time, and thus our results do not account for the net torque from multiple mergers.

Disk tilting has several other important ramifications for astrophysical observations and terrestrial dark matter experiments. First, disk tilting is likely to be an important consequence of the \emph{Gaia}-Sausage-Enceladus~(GSE) merger, one of the most significant events in the Milky Way's history~\citep{2018MNRAS.478..611B, 2018Natur.563...85H}. The consequence of the GSE for disk tilting was first explored by~\cite{2022MNRAS.513.1867D} using cosmological simulations. Our isolated $N$-body simulations allow us to confirm that a GSE-like merger may indeed cause significant and sustained disk tilting over a period of $\sim8$--10~Gyrs, with the tilting rate being  sensitive to the disk mass. These results suggest that the Sun's relative orientation may still be evolving in time today in response to the dynamical forces of the GSE debris. Given uncertainties in disk modeling, comparisons between simulations and data must therefore be made with care for observables that are sensitive to the Sun's relative location within the Milky Way host.

Second, disk tilting can affect the spatial and velocity distribution of DM near the Sun, which impacts direct detection experiments---see~\cite{2013RvMP...85.1561F,2014JPhG...41f3101R} for reviews.   This can have particularly drastic consequences if the tilting brings the stellar disk more (or less) into alignment with DM substructure in the halo. We find that significant enhancements of the high velocity tail of the DM's velocity distribution at the Solar position can occur for retrograde mergers with small inclination angles. This is especially relevant for DM models with large energy thresholds such as inelastic scenarios~\citep{Smith_2001}.

The paper is organized as follows.  Sec.~\ref{sec:disktilting} introduces an intuitive model of the key physical effects that determine the degree of disk tilting, including precession and angular momentum alignment. This is perhaps the most important section of the paper as it summarizes the key take-away conclusions regarding disk tilting dynamics.  Sec.~\ref{sec:methodology} presents the simulation details.  Sec.~\ref{sec:Results} presents the results of our simulation scans over satellite initial conditions and orbital properties, confirming the intuition from Sec.~\ref{sec:disktilting}. Sec.~\ref{sec:implications} discusses the implications of disk tilting for observations and experiments.  We conclude in Sec.~\ref{sec:conclusions}. Appendix~\ref{sec:Disk_Dynamics} presents a toy model for the dynamics of disk tilting during a satellite merger. Appendix~\ref{sec:supp_figures} includes supplementary figures and a table that provide the tilting rates for all satellite parameters considered in our study.

\section{Physics of Disk Tilting}
\label{sec:disktilting}

The goal of this section is to build intuition for how the DM that is accreted from a merging satellite impacts the tilting dynamics of a host galaxy's stellar disk, ahead of the dedicated simulation studies presented later in Sec.~\ref{sec:Results}. We suggest here that disk tilting can be understood in terms of the collective behavior of accreted DM (aDM) particles orbiting within the host's potential. The results obtained from the simple toy analysis discussed qualitatively in this section, and more quantitatively in App.~\ref{sec:Disk_Dynamics}, are corroborated by our full numerical study.

\begin{figure}
\includegraphics[width=0.48\textwidth]{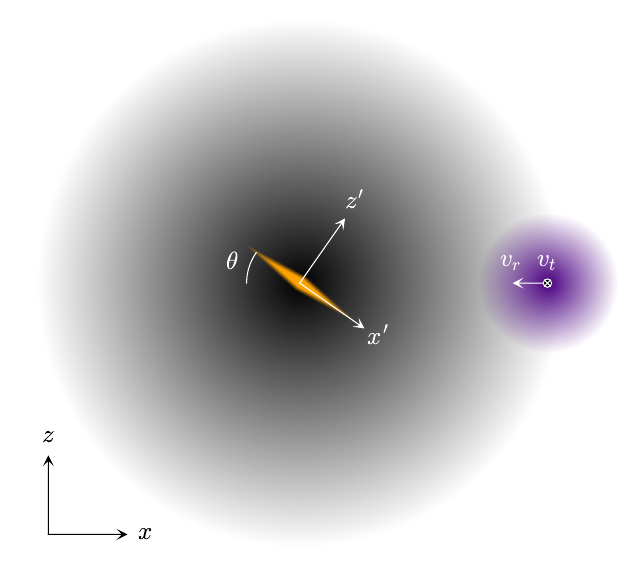}
\caption{Schematic diagram of the merger geometry. The satellite (purple) is initialized at some radius from the host center with orbital circularity $\eta \equiv \lvert v_t \rvert/\sqrt{v_r^2+v_t^2}$, where $v_r$ and $v_t$ are its radial and tangential velocities, respectively. The initial orbital plane of the satellite is denoted $x_{\rm orb} - y_{\rm orb}$, and the plane of the stellar disk (orange) is denoted $x_\star - y_\star$. The initial inclination angle between these planes is $\theta_{\rm inc}$. Orbits are always initialized such that the $\hat{y}_{\rm orb}$ and $\hat{y}_\star$ directions are aligned.}
\label{fig:geometry}
\end{figure}

We begin by understanding the phenomenon of disk tilting as a consequence of angular momentum conservation under a few simplifying assumptions: the angular momentum of any baryonic but non-stellar components (e.g.~gas) of the galaxy is negligible, and the DM halo is perfectly spherical and does not transfer angular momentum to or from the stellar disk or the satellite/aDM. In this case, the system composed of the stellar disk together with the progenitor satellite experiences a spherical gravitational potential, and its total angular momentum is conserved. A schematic diagram of this geometry is shown in Fig.~\ref{fig:geometry}. Initially, the satellite is a well-defined and self-bound object orbiting in a plane denoted as the $x_{\rm orb}-y_{\rm orb}$ plane. The orbiting, bound satellite can torque the disk and cause a brief period of tilting, but we focus on the period of tilting after accretion in the remainder of this section. The plane defined by the host's stellar disk is denoted as the $x_\star-y_\star$ plane, and the inclination angle between these planes is given by $\theta_{\rm inc}$ (defined to always be $\theta_{\rm inc} \leq 90^\circ$).

\begin{figure*}
    \centering
    \includegraphics[width=0.95\textwidth]{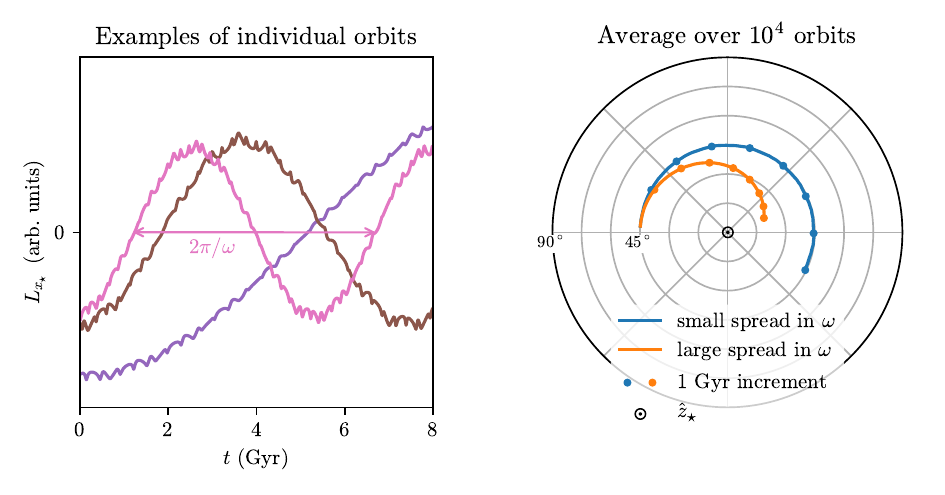}
    \caption{Results using a simple toy model for the angular momentum dynamics of the inner accreted dark matter (aDM), $\vec{L}_{\rm aDM}^{\rm inner}$. We assume that the aDM particles orbit a fixed Miyamoto Nagai stellar disk ($a=2$, $b=1$, $M_\star=2\times 10^{10}~M_\odot$) embedded within an NFW profile ($M_{\rm vir}=10^{11.75}~M_\odot$, $c_{\rm vir}=6.3$, $z=1$). The left panel shows three examples of the projection of individual aDM particles' angular momenta onto the $\hat{x}_\star$ axis. Each particle approximately precesses at some frequency $\omega$. The right panel is a Briggs plot that shows the evolution of the average angular momentum for an ensemble of $10^4$ aDM particles centered around $\hat{z}_\star$; a Briggs plot is a 2D polar plot that tracks the direction of the angular momentum vector. The blue curve corresponds to an ensemble with a very small spread in frequencies and thus the average angular momentum simply precesses around $\hat{z}_\star$. The orange curve corresponds to an ensemble with a large spread in frequencies and thus the average angular momentum precesses while also aligning with the $\hat{z}_\star$ axis. Each dot on the curves represents a $1$ Gyr increment in time.}
    \label{fig:Prec_Align_Toy}
\end{figure*}

After experiencing tidal forces from the host, the incoming satellite's mass becomes unbound and spreads throughout the halo---this is the aDM. Angular momentum conservation at all times $t$ enforces
\begin{equation}
    \vec{L}_\star(t) + \vec{L}_{\rm aDM}(t) = \vec{L}_{\rm tot}(0) \,,
\end{equation}
where $\vec{L}_\star(t)$ ($\vec{L}_{\rm aDM}(t)$) is the angular momentum of the stellar disk (aDM) at time $t$, and $\vec{L}_{\rm tot}(0)$ is the total angular momentum of the system at time $t=0$. After the entire mass of the satellite is disrupted, the aDM structure can be artificially divided into an inner and an outer component with angular momenta $\vec{L}^{\rm inner}_{\rm aDM}$ and $\vec{L}^{\rm outer}_{\rm aDM}$, respectively. By definition, the outer component experiences little torque from the inner regions, and thus its angular momentum is approximately constant in time. On the other hand, the inner component is much closer to the stellar disk and experiences a significant torque, which varies with time. 

\begin{figure*}
    \centering
    \includegraphics[width=\textwidth]{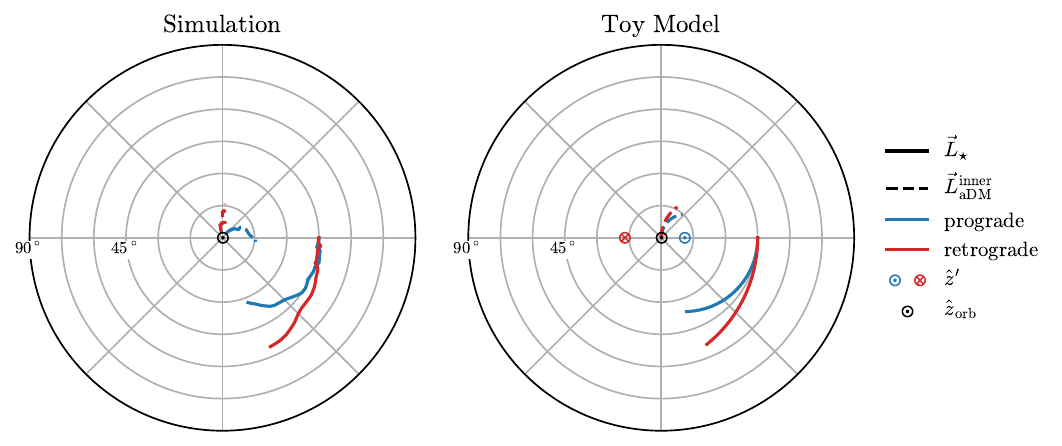}
    \caption{Dynamics of prograde versus retrograde orbits centered around $\hat{z}_{\rm orb}$. The left panel shows the evolution of $\vec{L}_\star$ (solid) and $\vec{L}_{\rm aDM}^{\rm inner}$ (dashed) from two high-resolution simulations of a $10^{11}~M_\odot$ satellite merging at $z=1$ with circularity $\eta=0.5$. Values of $\vec{L}_{\rm aDM}^{\rm inner}$ have been chosen heuristically to include all accreted dark matter~(aDM) particles within $12~\text{kpc}$ of the host's center. Blue curves correspond to prograde orbits while red curves correspond to retrograde orbits. In each case, the stellar and aDM angular momenta precess around the common axis, $\hat{z}'$. The right panel shows a toy model of these dynamics. For these curves, typical values of stellar and aDM angular momenta have been chosen (specifically $| \vec{L}_{\rm aDM}^{\rm inner}| =3 | \vec{L}_\star |$) and the result of precession around $\hat{z}'$ (as well as the direction of $\hat{z}'$) has been plotted for the two cases of opposite signs for $\vec{L}_\star$. For the prograde case, both $\hat{z}'$ and $\vec{L}_\star$ point out of the page while for the retrograde case, both vectors point into the page.}
    \label{fig:Pro_Retro_Toy}
\end{figure*}

Denoting the direction of the vector $(\vec{L}_{\rm tot}(0) - \vec{L}^{\rm outer}_{\rm aDM})$ as $\hat{z}'$, one finds that the angular momenta of the stellar disk and the inner aDM obey
\begin{equation}
(\vec{L}_\star)^T \simeq - (\vec{L}_{\rm aDM}^{\rm inner})^T \,,
\label{eq:L_T}
\end{equation}
where the $T$ superscript denotes the projection of these vectors onto the $x'-y'$ plane. The implication of Eq.~(\ref{eq:L_T}) is simply that $(\vec{L}_{\rm aDM}^{\rm inner})^T$ determines $(\vec{L}_\star)^T$ to a very good approximation, which provides a window into the stellar disk tilting dynamics. For example, as will be shown below, if one of these objects precesses around the $\hat{z}'$ axis, so does the other. A different phenomena occurs in the limit $(\vec{L}_{\rm aDM}^{\rm inner})^T \to 0$, in which case the stellar disk's $(\vec{L}_\star)^T$ also vanishes and the angular momenta of these two objects align (or anti-align) with each other. In fact, these two effects, namely precession of $\vec{L}_\star$ and $\vec{L}_{\rm aDM}^{\rm inner}$ around the common $\hat{z}'$ axis, and alignment of these vectors with the $\hat{z}'$ axis, are precisely those we observe when studying the results of the simulations.

A toy model that explains these dynamics by tracking the motion of aDM particles orbiting a \emph{fixed} stellar disk is given in App.~\ref{sec:Disk_Dynamics}; here, we simply summarize the main points. If indeed the stellar disk is fixed (e.g.~as would be the case if  $|\vec{L}_\star| \gg |\vec{L}_{\rm aDM}^{\rm inner}|$), then each aDM particle's angular momentum approximately precesses around that of the stellar disk. Figure~\ref{fig:Prec_Align_Toy} (left panel) shows some numerical examples of the projection of angular momentum onto the $\hat{x}_\star$ axis (which in the limit $|\vec{L}_\star| \gg |\vec{L}_{\rm aDM}^{\rm inner}|$ is just equal to the $\hat{x}'$ axis) for particles with varying initial conditions orbiting in the potential of a Miyamoto-Nagai stellar disk~\citep{1975PASJ...27..533M} and a Navarro-Frenk-White~(NFW) host halo~\citep{Navarro_1996}. It is clear that each particle's angular momentum approximately precesses around that of the stellar disk with some frequency $\omega$.

If all orbiting aDM particles have similar values of $\omega$ and their phases are coherent, this just corresponds to an overall precession motion. However, over time, these phases begin to de-cohere and eventually the average angular momentum of all orbiting aDM particles is simply aligned or anti-aligned with that of the stellar disk. Such behavior can be seen in the right panel of Fig.~\ref{fig:Prec_Align_Toy}.\footnote{This is an example of a Briggs plot~\citep{1990ApJ...352...15B}, a 2D polar plot that maps out the angular coordinates of a vector, ignoring its magnitude. Note that in many of the Briggs plots presented in this study, there is also no distinction between vectors that point into versus out of the page.} This figure provides two examples of the dynamics of $\vec{L}_{\rm aDM}^{\rm inner}$ (from which $\vec{L}_\star$ can be inferred) centered around $\hat{z}_\star$, for large ensembles of aDM particles. Each curve is calculated by numerically injecting many aDM particles into orbits and averaging over their angular momenta at each timestep. The blue curve shows the result for a very coherent set of orbiting aDM particles (similar values of $\omega$), while the orange curve shows the result for a very incoherent set. Each dot on the curves represents a $1$ Gyr increment in time. In the former case, $\vec{L}_{\rm aDM}^{\rm inner}$ simply precesses, as evidenced by the fact that the angular momentum remains at a fixed angular distance from the origin. In the latter case, $\vec{L}_{\rm aDM}^{\rm inner}$ precesses and aligns with the angular momentum of the stellar disk, as evidenced by the fact that the angular momentum drifts towards the origin, $\hat{z}_\star$. These curves can be compared to the analytical results of App.~\ref{sec:Disk_Dynamics}.

\begin{table*}
    \centering
    \begin{tabular}{lrllllll}
        
\hline\hline
        
Component & Parameter & Host & S5 & S4  & S3  & S2 & S1 \\

\hline

DM Halo     & $\log_{10} M_\text{vir}/M_\odot$   & 11.85 & 11.00 & 10.75 & 10.50 & 10.25 & 10.00 \\
          & $R_\text{vir}$ (kpc)                     & 136 & 71  & 59 & 49 & 40 & 33 \\
            & $c_\text{vir}$                     & 6.3  & 7.2  & 7.5 & 7.9 & 8.2 & 8.6 \\
            & Hernquist $a$ (kpc)            & 31  & 15  & 12  & 9.6 & 7.7 & 6.2  \\
\hline
Stellar Disk  & $\log_{10} M_\star/M_\odot$ & 10.2 & 8.5  & 8.0 & 7.5  & 7.0 & 6.5 \\
 & $R_\star$ (kpc)             & 2.3  & 1.0  & 0.85 & 0.68  & 0.55 & 0.44 \\
              & $Z_\star$ (kpc)             & 0.90  & 0.43  & 0.34 & 0.27  & 0.22 & 0.18 \\

\hline\hline

    \end{tabular}
    \caption{ Initial conditions for our suite of simulations. All galaxies are constructed at $z=1$ according to the procedure discussed in Sec.~\ref{sec:initial_conditions}. The table provides the virial mass $M_\text{vir}$, radius $R_\text{vir}$, and concentration $c_\text{vir}$, for the host as well as properties of the five satellite types~(S1-S5). The dark matter halos are specified by NFW parameters and converted by \texttt{GALIC} into Hernquist parameters that give a similar inner density profile. The Hernquist scale radius, $a$, is also provided for convenience.  Additionally, each galaxy has an assigned exponential stellar disk, and we provide its mass $M_\star$, scale radius $R_\star$, and scale height $Z_\star$.}
    \label{tab:galaxy_ic}
\end{table*}

In the more realistic scenarios observed in our simulations, many DM particles will be stripped from their satellite galaxy over a short period of time when the satellite is close to pericenter. Since the particles are stripped over a relatively short time, the initial orbiting aDM particles have coherent frequencies corresponding to an initial precession motion. Over time, the particles begin to de-cohere and the dynamics begins to cause alignment. Importantly, since these effects depend on the orbital dynamics of aDM particles, there is a non-trivial dependence of the tilting rates on the inclination angle $\theta_{\rm inc}$. Specifically, when $\theta_{\rm inc} = 0^\circ$ or $90^\circ$, aDM particles orbit in a single plane and their angular momenta remain constant. Thus, the tilting rate approaches zero for both these cases and is maximal at some intermediate value of $\theta_{\rm inc}$.

The $\hat{z}'$ direction itself is set by the initial angular momentum of the bound satellite, the initial angular momentum of the stellar disk, and the value of $\vec{L}_{\rm aDM}^{\rm outer}$. This provides additional intuition regarding the possible motion of the stellar disk. For example, for any incoming satellite, the difference between a prograde and retrograde orbit corresponds to opposite directions of $\vec{L}_\star$. This feeds into the direction of $\hat{z}'$, which sets the axis around which any precession can occur. Thus, a stellar disk that precesses as a reaction to torques from an incoming satellite will do so around one of two possible axes depending on whether the stellar disk is  prograde or retrograde with respect to the satellite's orbit.

This difference between prograde and retrograde precession is demonstrated explicitly in Fig.~\ref{fig:Pro_Retro_Toy}.
The left panel shows results from two of our simulations~(which will be described in detail in Sec.~\ref{sec:methodology}). The figure shows the evolution of $\vec{L}_\star$ (the angular momentum of all host star particles) and $\vec{L}_{\rm aDM}^{\rm inner}$ (chosen heuristically to be all aDM particles within $12~\text{kpc}$ of the host center) for a given set of initial satellite mass and orbital parameters, and for two opposite values of initial $\vec{L}_\star$---prograde in blue and retrograde in red, centered around $\hat{z}_{\rm orb}$, which can be calculated from the initial orbit of the bound satellite. In each case, $\vec{L}_\star$ and $\vec{L}_{\rm aDM}^{\rm inner}$ approximately precess around a common axis that varies for the prograde versus retrograde orbits. The right panel shows a toy model that demonstrates the same behavior. Here, typical values of stellar and aDM angular momenta have been chosen, and the result of precession around $\hat{z}'$ (as well as the explicit direction of $\hat{z}'$) has been plotted for the two cases with opposite signs of $\vec{L}_\star$. Note that for the prograde case, both $\hat{z}'$ and $\vec{L}_\star$ point out of the page, while for the retrograde case both vectors point into the page. This toy model should not be regarded as accurately modeling the observed behavior, but rather simply as a proof of concept that the observed differences in the dynamics of $\vec{L}_\star$ between prograde and retrograde orbits can be traced back to the different $\hat{z}'$ directions.

\section{Methodology} 
\label{sec:methodology}

This section describes the numerical framework utilized to perform the simulations of isolated galaxy mergers presented in this work.  We are primarily interested in tracking a given satellite's evolution as it falls into a Milky Way--like host, in order to understand its impact on the disk tilting rate. Our goal is to explore how  disk tilting depends on the initial conditions: to this end, we vary over parameters such as the satellite mass, the circularity and inclination of its orbit, and the infall direction (prograde/retrograde). In all simulations, we set the initial $z_{\rm orb} - z_\star$ plane to be perpendicular to the orbital plane of the satellite (the $\hat{y}_{\rm orb}$ and $\hat{y}_\star$ directions are aligned), since the precessing motion of the stellar disk typically scans over this additional angle. Section~\ref{sec:initial_conditions} describes the initial conditions for the host and satellite systems, followed by Sec.~\ref{sec:simulations} which describes the configuration of the $N$-body simulations in detail.

\subsection{Initial Conditions}
\label{sec:initial_conditions}

Each galaxy is modeled using an NFW density profile, which is set by two free parameters arbitrarily chosen to be the halo virial mass $M_\text{vir}$, and the concentration $c_\text{vir}$. We assume a $\Lambda$CDM cosmology with $h=0.7$, $\Omega_m=0.3$, and $\Omega_\Lambda=0.7$. The halo's virial mass $M_\text{vir}$ is related to its virial radius $R_\text{vir}$ through 
\begin{equation}
 M_\text{vir} = \frac{R_\text{vir} \Delta_c H^2}{2 G} \, ,
\end{equation}
where $H$ is Hubble's constant, $G$ is Newton's gravitational constant, and $\Delta_c$ is the overdensity parameter as defined in \citet{Bryan_1998}. For $z=1~(2)$, we take $\Delta_c=157~(171)$.  The concentration-mass relation from \citet{Dutton_2014} is used to estimate the halo concentration for each galaxy.  Table~\ref{tab:galaxy_ic} summarizes the initial conditions in detail.

We consider five different satellite masses (S1-S5): $\log_{10} M_{\rm vir}/M_\odot= 10, 10.25, 10.5, 10.75, 11$ and use a host mass $M_{\rm vir} = 7\times 10^{11} M_\odot$ based on the mean $z=1$ mass for Milky Way-like merger trees. These trees are generated using the \texttt{SatGen}\footnote{\href{https://github.com/shergreen/SatGen}{https://github.com/shergreen/SatGen}} code~\citep{2021MNRAS.502..621J}, selecting galaxies with a $z=0$ mass between $10^{12.1}$ and $10^{12.3} M_\odot$. The code uses the extended Press-Schechter formalism \citep{Lacey_Cole_1993MNRAS.262..627L} for hierarchical structure formation incorporating modifications \citep{Parkinson_2007} that result in better agreement with cosmological simulations.

All galaxies include an exponential stellar disk with scale radius $R_\star$ and scale height $Z_\star$. The stellar-to-halo mass relation from \citet{Rodr_guez_Puebla_2017} is used to determine the mass of the disk for a given $M_{\rm vir}$.  The scale radius is set using the relation from \citet{Jiang_2019}:
\begin{equation}
    R_\star = 0.02 \left(\frac{c_\text{vir}}{10} \right)^{-0.7} R_\text{vir} \,.
\end{equation}
Additionally, the scale height of the disk is set to 
\begin{equation}
Z_\star = 0.4 \,R_\star \,,
\end{equation}
and the disk velocity dispersion ratio is set to $\sigma_{R_\star}/\sigma_{z_\star} =4$. The requirement on $Z_\star$ results in a scale height of 900~pc for the host galaxy at $z=0$, comparable to the thick-disk expectation for the Milky Way~\citep{2016ARA&A..54..529B}. However, the rotational disk velocities in our simulations at $z=0$ are systematically lower than observed values. This is likely due to the fact that we do not include a bulge nor do we model disk growth with time. We do not rescale the velocity values as was done in~\cite{naidu2021reconstructing} because the tilting rate is not sensitive to this choice. Lastly, we note that we have evolved the disks in isolation and explicitly verified that they remain stable with no bar formation over the simulation timescale, consistent with the fact that the Toomre parameter~\citep{1964ApJ...139.1217T} everywhere on the disk is $Q > 2.5$.  

We use \texttt{GALIC}\footnote{\href{https://www.h-its.org/2014/11/05/galic-code/}{https://www.h-its.org/2014/11/05/galic-code/}}~\citep{2014MNRAS.444...62Y} to generate near-equilibrium positions and velocities for the stellar and DM particles in the satellite and host galaxies.  \texttt{GALIC} initializes the halo using the Hernquist profile~\citep{1990ApJ...356..359H}, and we provide the corresponding Hernquist scale radius in Tab.~\ref{tab:galaxy_ic} for convenience. Adiabatic contraction~\citep{1986ApJ...301...27B, Gnedin:2004cx} is not incorporated in our initial conditions, nor do we observe significant deviations from halo isotropy over the course of our simulations. Note that we tune the spin parameter $\lambda$ to reproduce the correct disk scale radius $R_\star$. 

For each merger configuration, the satellite galaxy is positioned at the virial radius of the host at an inclination $\theta_{\rm inc}$ from the host disk plane (see Fig.~\ref{fig:geometry}). The prograde (retrograde) merger direction specifies the infall direction along (counter to) the spin of the host disk. We set the spin direction of the satellite disk to match that of the host disk. The initial energy of the satellite is set equal to that of a circular orbit at $R_\text{vir}$. The orbit circularity $\eta$ is defined as
\begin{equation}
    \eta \equiv \lvert v_t \rvert / \sqrt{v_t^2+v_r^2} \, ,
\end{equation}
where $v_t$ is the transverse velocity and $v_r$ is the radial velocity. 
Cosmological simulations suggest that the most common circularity for mergers such as those studied in this work is around $\eta=0.5$, with a wide spread \citep{Wetzel_2010}. We note that extreme values ($\eta=0$ and $\eta=1$) are unlikely.

\begin{figure*}
    \centering
    \includegraphics[width=\textwidth]{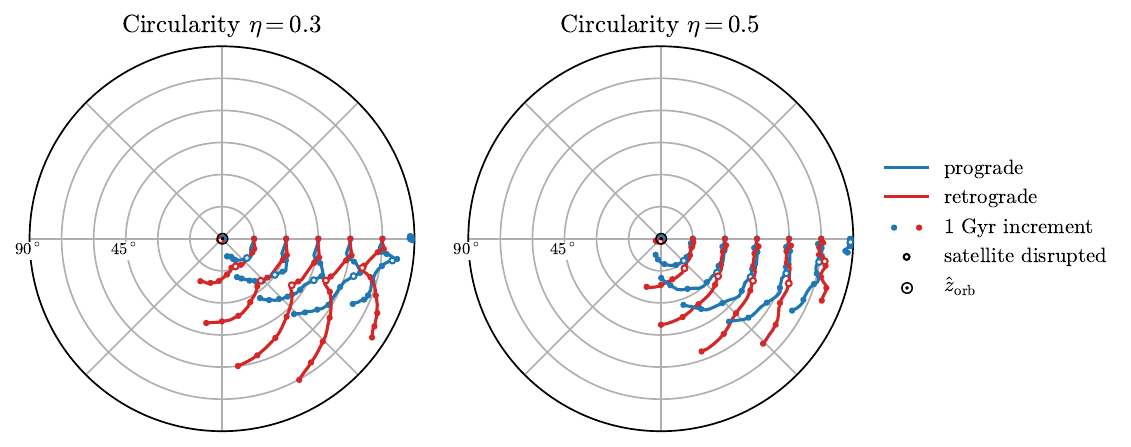}
    \caption{Evolution of the host stellar disk's angular momentum for the case of a $10^{11} M_\odot$ satellite initialized at $z=1$ at the host virial radius, with inclination angle $\theta_{\rm inc}$ on a prograde~(blue) or retrograde~(red) orbit. Two circularities are considered: $\eta = 0.3$~\emph{(left panel)} and $\eta = 0.5$~\emph{(right panel)}. Solid markers are placed at 1 Gyr increments; the open circle denotes the point in time when the satellite becomes completely disrupted. In some cases, significant disk tilting occurs before this point due to torques from the orbiting bound satellite core, but this is difficult to see in the low-resolution simulations. See Fig.~\ref{fig:app_highres_polar} for selected high-resolution simulations. For these examples, the precession axis is approximately equal to the direction of the satellite's initial orbital angular momentum, $\hat{z}' \simeq \hat{z}_{\rm orb}$. Precession is identified as motion about $\hat{z}_{\rm orb}$ at fixed angular distance. However, depending on the orbital parameters of the satellite, the angular momentum of the host disk can either be pushed into or out of alignment with the angular momentum of the accreted DM.}
    \label{fig:polar_plot}
\end{figure*}

\subsection{N-Body Simulations} 
\label{sec:simulations}

We use the GIZMO\footnote{\href{https://bitbucket.org/phopkins/gizmo-public/src/master/}{https://bitbucket.org/phopkins/gizmo-public/src/master/}} package~\citep{Hopkins_2015} to model the $N$-body gravitational interactions of the stars and DM particles involved in the satellite mergers.  Our baseline simulations assume a collision-less particle mass $m_p=10^{5.5} M_\odot$ for both the DM and stars.  We use $\epsilon_\text{acc}$ from  \cite{Power_2003} to obtain the minimum softening length of 85~pc. We consider eleven evenly-spaced circularities in the range $\eta \in [0, 1]$, inclinations of $\theta_{\rm inc} = [0^\circ,  15^\circ, 30^\circ, 45^\circ, 60^\circ, 75^\circ, \text{ and } 90^\circ]$, prograde and retrograde infall directions, and satellite masses $\log_{10}M_{\rm sat}/M_\odot=[10.0,10.25,10.5,10.75, \text{ and } 11.0]$. This implies a scan of 685 combinations of satellite mass, circularity, inclination, and infall direction when eliminating redundant parameter combinations.

At the resolution used for the baseline runs, we find that the stellar disk grows 8\% thicker over the course of the simulation even without a merger, as measured by the half-mass height $z_{1/2}$ (defined such that half the disk mass is contained within $[-z_{1/2},z_{1/2}]$). This is likely due to relaxation effects, see e.g.~\citet{Sellwood_2013}. To better understand the effects of the merger on the stellar disk, we also simulated a handful of qualitatively interesting cases with higher resolution $m_p=10^{4.5} M_\odot$ and softening length of 27~pc. These runs have comparable resolution to other state-of-the-art $N$-body simulations performed for the Sagittarius~\citep{2018MNRAS.481..286L}, Large Magellanic Cloud~\citep{Garavito_Camargo_2019}, and GSE~\citep{naidu2021reconstructing} mergers. Table ~\ref{tab:app_highres} lists the parameters for the simulations run at high resolution.

\section{Results}
\label{sec:Results}

Section~\ref{sec:disktilting} outlined the primary physical effects that drive the angular momentum evolution of a host galaxy's stellar disk. The arguments presented there were based on a simplified description of the problem. Given the setup described in the previous section, we are now equipped to understand how these results generalize by analyzing the output of our semi-realistic simulations. As is shown below, the results from simulations qualitatively follow the intuition developed in Sec.~\ref{sec:disktilting}.

The behavior of the host disk's total angular momentum ($\vec{L}_\star$) for the specific case of a $10^{11}~M_\odot$ satellite 
is summarized in Fig.~\ref{fig:polar_plot}. We provide results for initial inclination angles $\theta_{\rm inc} = 0^\circ, 15^\circ, 30^\circ, 45^\circ, 60^\circ, 75^\circ,$ and $90^\circ$ in both panels. The left~(right) panel of the figure shows the results for circularity $\eta = 0.3$~(0.5).  The blue curves correspond to prograde orbits, while the red curves correspond to retrograde orbits. For each configuration, the angular momentum evolution is tracked for a period of 8~Gyrs following infall; the solid dots along the curves in Fig.~\ref{fig:polar_plot} indicate 1~Gyr increments in the simulation. The plots are centered in the $\hat{z}_{\rm orb}$ direction.

\begin{figure*}
    \centering
    \includegraphics[width=0.98\textwidth]{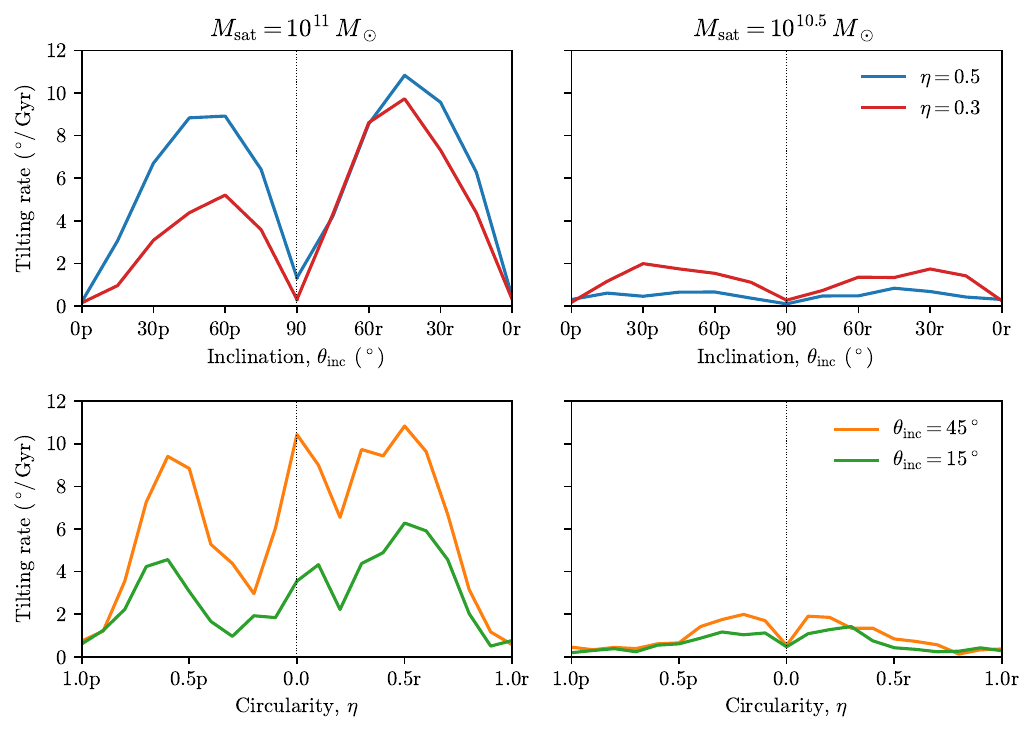}
      \caption{The tilting rate of the total angular momentum of the host galaxy's stellar disk. The rates are evaluated over the period of 6--8~Gyr from the start of the simulation under the assumption of a constant tilting rate in this interval. Left versus right panels correspond to two satellite masses, $10^{11}$ and $10^{10.5}~M_\odot$, respectively. Top panels show the tilting rate as a function of the satellite's orbital inclination angle $\theta_{\rm inc}$ for two values of circularity, $\eta = 0.3$ and $0.5$. Bottom panels show the rate as a function of $\eta$ for two values of $\theta_{\rm inc} = 15^\circ$ and $45^\circ$. These examples demonstrate that disk tilting rates can be substantial and depend sensitively on the details of a satellite's orbital parameters and mass. In general, rates are larger for higher satellite masses. For a given mass and circularity, the tilting rate is maximal for intermediate inclinations of $\theta_{\rm inc}\sim45^\circ$ and essentially vanishes for $\theta_{\rm inc} = 0^\circ$ and $90^\circ$. For a given mass and inclination angle, the rate is maximal at specific values of $\eta$. Lower mass satellites have peak tilting rates at smaller values of $\eta$, explaining the behavior in the top two panels.} Tilting rates are also typically larger for retrograde~(r) orbits compared to prograde~(p) ones. For completeness, results from all simulations are presented in App. \ref{sec:supp_figures}.
    \label{fig:tip_precession}
\end{figure*}

\begin{figure}
    \centering
    \vspace{5pt}
    \includegraphics[width=0.5\textwidth]{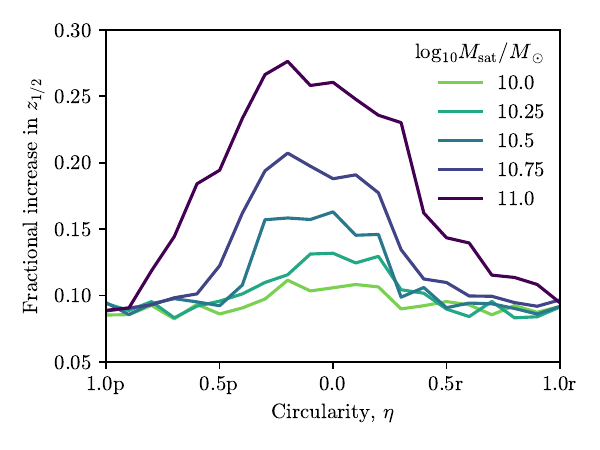}
    \caption{Fractional increase in half-mass height of the host stellar disk, $z_{1/2}$, from the initial to final time ($t=8~\mathrm{Gyr}$) of the simulation.  Results are shown for all satellite masses and circularities that were simulated, assuming a fixed orbital inclination angle of $\theta_{\rm inc} = 45^\circ$. Disk thickening is most pronounced for large satellite masses and small circularities. Dependence on inclination is less pronounced, as shown in App.~\ref{sec:supp_figures}.}
    \label{fig:disk_heating}
\end{figure}

We define the time of complete satellite disruption $t_\text{d}$ as the time at which all particles have become unbound from the satellite core. We follow the procedure from \textsc{Amiga}'s halo finder \citep{2009ApJS..182..608K} to identify unbound particles. In short, this procedure assumes spherical symmetry for the satellite core to calculate the potential using all satellite particles. Then, particles with a velocity greater than their escape velocity are removed, and the potential is recalculated. This process is repeated until we are left with a set of self-bound satellite particles. We use the center of mass of the initially innermost 1\% of satellite particles as a proxy for the core's location.

The open circle in Fig.~\ref{fig:polar_plot} indicates $t_\text{d}$. When $t \lesssim t_\text{d}$, the host's stellar disk is dominantly responding to the gravitational forces of the \emph{bound} satellite. Once the satellite is completely disrupted ($t \gtrsim t_{\rm d}$), so that the accreted DM is dispersed throughout the host, it still provides a significant torque on the stellar disk. At this point, the dynamics described in Sec.~\ref{sec:disktilting} become relevant. In particular, the stellar disk's angular momentum (as well as that of the inner aDM) begins to precess around the $\hat{z}'$ axis. Disk precession is apparent in the plots---the red/blue curves move in the angular direction, i.e., the right-to-left direction across radial spokes.

For a given set of satellite orbital initial conditions, the direction of $\hat{z}'$ strongly depends on whether the stellar disk is spinning in the prograde or retrograde direction. Additionally, given enough time, decoherence of the phases of angular momenta of orbiting aDM particles can cause alignment of the stellar disk's angular momentum vector with the $\hat{z}'$ direction. In the Briggs plot of Fig.~\ref{fig:polar_plot}, the combination of these two effects is observed as shifts of the red/blue curves from one gray semi-circle to another.

Fig.~\ref{fig:tip_precession} explicitly shows the overall tilting rate of the host's stellar disk, which is calculated by finding the angular distance between $\vec{L}_\star$ over the period 6--8~Gyrs and assuming a constant tilting rate between these two times.\footnote{We have verified that the results do not change qualitatively if the rate is evaluated over any $2$~Gyr period beyond $5$~Gyrs from infall.} The result is shown for a satellite of mass $10^{11}~M_\odot$~(left column) and $10^{10.5}~M_\odot$~(right column). The top panels of Fig.~\ref{fig:tip_precession} show the tilting rate as a function of inclination angle for two circularities, $\eta = 0.3$ and $0.5$, both prograde and retrograde (note that the top left panel corresponds to the same simulations as shown in Fig.~\ref{fig:polar_plot}). The bottom panels of Fig.~\ref{fig:tip_precession}  show the tilting rate as a function of circularity for two values of inclination angle, $\theta_{\rm inc} = 15^\circ$ and $45^\circ$, both prograde and retrograde.

Changes to the stellar disk's angular momentum direction are much larger for $10^{11}~M_\odot$ than for $10^{10.5}~M_\odot$, the reason being that larger satellite masses corresponds to larger values of $\vec{L}_{\rm aDM}$, which also corresponds to more torque experienced by the stellar disk (and also to $\hat{z}'$ being more displaced from $\hat{z}_\star$). From the top panels, it is evident that the tilting rate is minimal for $\theta_{\rm inc} \sim 0^\circ, 90^\circ$ and maximal for $\theta_{\rm inc} \sim 45^\circ$; this point was explained in Sec.~\ref{sec:disktilting}. From the bottom panels, one can note a preference for larger tilting rates at specific values of $\eta$. For example, the bottom left panel shows larger rates for $\eta \sim 0$ and $0.6$. At $\eta = 0$ (maximally radial orbits), the dynamics are such that the stellar disk in fact precesses approximately around the $\hat{x}_{\rm orb}$ direction with a sizable overall tilting rate. At $\eta \sim 0.6$, precession around $\hat{z}'$ also causes a large tilting rate. Approaching $\eta \sim 1$ (circular orbits), much less DM is stripped from the bound satellite and so disk tilting is not a large effect. 

A final feature observed in Fig.~\ref{fig:tip_precession} is that tilting rates are typically faster for retrograde orbits compared to prograde orbits. This is an artifact of the angle between $\vec{L}_\star$ and $\hat{z}'$ typically being larger for retrograde orbits than for prograde orbits. Therefore, when $\vec{L}_\star$ precesses around $\hat{z}'$, it sweeps over angles at a higher rate for retrograde orbits than for prograde ones.

Interestingly, the tilting rates shown in Fig.~\ref{fig:tip_precession} are significant enough to fall within the \emph{Gaia} detection range, approximately $> 0.28\,^\circ /$Gyr~\citep{Perryman:2014gka}. Tilting rates from our entire suite of simulations are presented in App.~\ref{sec:supp_figures}. As can be seen from those results, or from the subset shown in Fig.~\ref{fig:tip_precession}, tilting rates are potentially measurable even for low-mass satellites and for a wide range of the orbital parameters $\theta_{\rm inc}$~and~$\eta$. This motivates follow-up simulation studies with improved disk and halo modeling in order to make reliable predictions for observational data in the Milky Way.

In addition to characterizing the disk tilting properties, it is important to understand whether the disk morphology is reasonably well preserved in the process. To this end, we evaluate the co-rotation parameter $\kappa_{\rm co}$, which captures the fraction of co-rotating particles in the stellar disk~\citep{2010MNRAS.409.1541S}.  It is defined as 
\begin{equation}
\kappa_{\rm co} = \frac{1}{K_{\rm tot}} \sum_i \frac{1}{2} \, m_i \, v_{\theta, i}^2\, ,
\end{equation}
where $m_i$ is the mass of the $i^{\rm th}$ stellar particle in the region with $r<30$~kpc and with $v_\theta > 0$.  $K_{\rm tot}$ is the kinetic energy of \emph{all} particles in this spatial region.  As has been shown in~\cite{2017MNRAS.472L..45C}, $\kappa_{\rm co}$ is a good proxy for the morphology of simulated galaxies, with values $\gtrsim 0.4$ indicating a disk-like morphology for the stellar distribution.  We find that $\kappa_{\rm co}$ does not change significantly with time, decreasing from 0.88 at early times to 0.85 for a few orbital configurations, but remaining about 0.87 for the majority of parameter space.
In addition to the overall disk morphology, we also explore the impact of the satellite mergers on the half-mass scale height, $z_{1/2}$.  Fig.~\ref{fig:disk_heating} shows the fractional change of $z_{1/2}$ over the course of the simulation, for a range of satellite masses and circularities, but fixed inclination angle of $\theta_{\rm inc} = 45^\circ$. Larger satellite masses and smaller circularities tend to thicken the disk more substantially. The largest effects are an increase of the scale height by about $30\%$. Given differences in the simulation initial conditions,  it is challenging to perform a direct comparison with other studies of isolated galaxy mergers. However, our finding that the disk is not destroyed by these encounters but may be thickened by them is consistent with, e.g., \cite{2008ApJ...688..757H,villalobos2008simulations,2008MNRAS.389.1041R,2009ApJ...703.2275P, 2008ApJ...688..757H}. Some of the mergers considered in this study may be excluded by other morphological or kinematic changes they induce to the stellar disk beyond tilting; a full exploration of this is beyond the scope of the present work.

Lastly, we have explored whether the particle resolution of the simulations affects the disk tilting results by performing dedicated high-resolution runs for the specific case of a $10^{11}~M_\odot$ satellite galaxy on a prograde/retrograde orbit with inclination $\theta_{\rm inc} = 45^\circ$ and circularity $\eta = 0.5$.  The particle mass in these runs is $10^{4.5}~M_\odot$, compared to the lower-resolution particle mass of $10^{5.5}~M_\odot$ that was used for all figures in this section. The high-resolution results are provided in Fig.~\ref{fig:Pro_Retro_Toy} and can be directly compared to the corresponding curves for the lower-resolution runs in Fig.~\ref{fig:polar_plot}.  Reassuringly, the angular momentum evolution of the host disk is indistinguishable between these cases (see also Tab.~\ref{tab:app_highres}).

\section{Discussion}
\label{sec:implications}

Before concluding, we take the opportunity to comment on potential implications of disk tilting, focusing specifically on the GSE merger (Sec.~\ref{sec:gse}) and direct detection experiments (Sec.~\ref{sec:direct_detection}). By design, the following presentation is illustrative in nature, focusing on the relevant lessons for each application. More detailed predictions, especially with regards to the GSE merger example, will benefit from improved modeling of the host's halo and disk in the simulation framework.

\subsection{Lessons for the GSE Merger}
\label{sec:gse}

\begin{figure*}
    \centering
    \includegraphics[width=\textwidth]{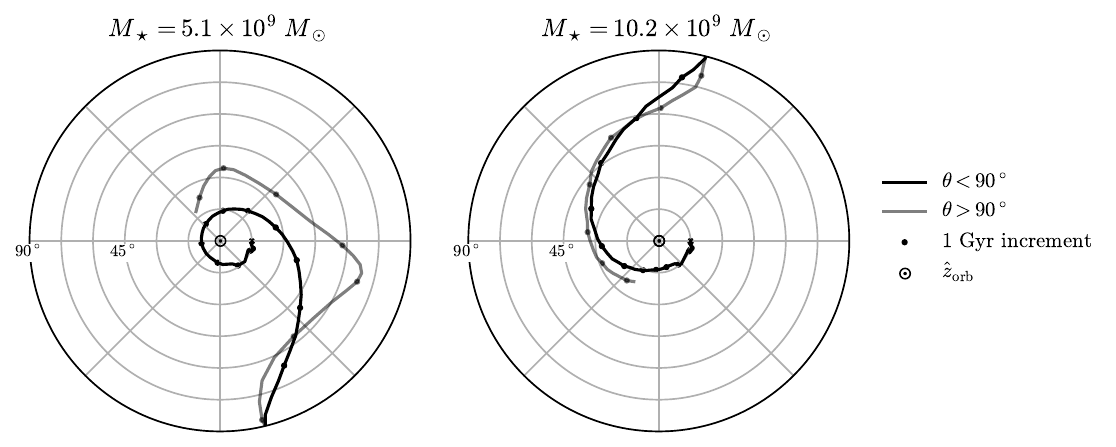}
    \caption{The results of an idealized simulation of the \emph{Gaia}-Sausage-Enceladus~(GSE) merger are presented here. We simulate a $M_{\rm vir} = 2.1\times10^{11}~M_\odot$ satellite falling into a Milky Way-like host at redshift $z=2$ with circularity $\eta = 0.5$ and retrograde inclination of $\theta_{\rm inc} = 15^\circ$. The left Briggs plot assumes a host disk mass of $M_{\rm \star} =5.1\times10^9~M_\odot$, while the right shows the results for double the host disk mass. The host disk tilts substantially in response to the merger, eventually flipping over. Changes to the disk mass or the integration time of the simulation can alter the disk tilting rate and---consequently---the relative position of the sun to the GSE tidal debris at $z=0$. As a result, an observer at the solar position would observe  pile-up of the DM and stellar debris at different locations on the sky in these different scenarios.}
    \label{fig:gse_merger}
\end{figure*}

Due to recent advancements in astrometric and spectroscopic observations, a clearer picture is emerging of the Milky Way's current dynamical state and historical evolution~\citep{2020ARA&A..58..205H}.  In particular, data from the~\emph{Gaia} satellite has been used to identify the remnants of what is likely to be the Galaxy's most recent major merger~\citep{2018MNRAS.478..611B, 2018Natur.563...85H}.  With an estimated total stellar mass of $M_{*} \sim 10^{8-9} M_{\odot}$ at an infall time of $\sim 8$--10~Gyr ago, the GSE merger likely contributed a large fraction of the stellar halo within $\sim 25$--30~kpc of the Galactic Center~\citep{deason_apocenter_2018,Haywood_2018,Myeong_2018,necib_under_2019, koppelman_one_2018, 2019MNRAS.487L..47V, 2019NatAs...3..932G, 2019A&A...630L...4M,lancaster_halos_2019,2019A&A...632A...4D,2019MNRAS.488.1235M, mackereth_origin_2019,bird_anisotropy_2019,necib_under_2019, 10.1093/mnras/staa047,bonaca_timing_2020,naidu_evidence_2020, 2020MNRAS.493..847F, koppelman_massive_2020,das_ages_2020,yuan_low-mass_2020, 2020MNRAS.498.2472K,feuillet_selecting_2021,Hasselquist_2021,iorio_chemo-kinematics_2021,carollo_nature_2021,gudin_r-process_2021,limberg_targeting_2021,2022ApJ...932L..16D,2021NatAs...5..640M,2022MNRAS.510.2407B}.

We have performed dedicated high-resolution simulations of a GSE-like merger to study the resulting degree of disk tilting. Following the same procedure as in Sec.~\ref{sec:initial_conditions}, except this time initializing the host galaxy at $z=2$, we generate a Milky Way analogue with halo mass $M_\text{vir}=5.3\times10^{11} M_\odot$, virial radius $R_{\rm vir} =85~\text{kpc}$, and concentration $c_{\rm vir} = 4.6 $. The host's stellar disk has mass $M_\star=5.1\times 10^9~M_\odot$, scale radius $R_\star =1.7~\text{kpc}$, and scale height $Z_\star =0.70~\text{kpc}$. (We have also simulated a GSE-like merger event with double the host stellar disk mass; see details below.) The GSE-analogue is modeled following the best-fit parameters from~\cite{naidu2021reconstructing}, with a virial mass $M_{\rm vir} =2.1\times10^{11}~M_\odot$, inclination $\theta_{\rm inc} = 15^\circ$, and retrograde circularity $\eta = 0.5$, and its orbit is evolved over a period of 10~Gyrs. We have verified that the resulting velocity distributions for the accreted GSE stars are similar to observational results, reproducing the ``sausage-like'' distribution in the radial and azimuthal plane of Galactocentric velocities.\footnote{
Note that our results will not be an exact replication of the dedicated GSE simulations of~\cite{naidu2021reconstructing} given differences in our respective setups, including the fact that we do not include a bulge for the host and use slightly different prescriptions to define the host and satellite initial conditions.}

The Briggs plot in the left panel of Fig.~\ref{fig:gse_merger} demonstrates how the host disk angular momentum evolves over the course of GSE infall for the case of $M_\star=5.1\times 10^9~M_\odot$. Initially, there is a clockwise precession approximately about the $\hat{z}_{\rm orb}$ direction, but then the disk angular momentum starts to rotate to anti-align itself with the angular momentum of the GSE debris. After about 7~Gyrs, the host disk flips over entirely, and the precession appears to change direction (although this statement depends on the choice of reference frame). We find initial tilting rates of $\sim20\,^\circ\!/\mathrm{Gyr}$ at $t=4~\text{Gyr}$, increasing to a maximum disk tilting rate of $\sim90\,^\circ\!/\text{Gyr}$ at $t=8~\text{Gyr}$, and then slowing to $\sim40\,^\circ\!/\text{Gyr}$ at $t=10~\text{Gyr}$ (near $z=0$). These values are averaged over the $0.12~\text{Gyr}$ simulation timestep. At $z=0$ in the simulation, the host disk is still tilting in response to the merger. Importantly, we find that slight changes in the integration time (the time in the simulation defined as $z=0$) can correspond to significantly different orientations of the host's stellar disk.

We also check the extent to which the host disk is disrupted by the merger using the same approach as for our grid scans. The corotation parameter $\kappa_{\rm co}$ decreases from an initial value of 0.90 to 0.76, a large difference but still disk-like. The half-mass scale height $z_{1/2}$ increases by about 70\%. These more dramatic changes are expected because of the larger mass of GSE compared to the satellites explored in our grid scans.

We also find that the evolution of the stellar disk in response to the GSE is sensitive to assumptions made about the disk modeling. To illustrate this point, we repeat the simulation, but double the mass of the host's stellar disk. This simulation provides an indication of how our results might change for the more realistic scenario of a stellar disk which grows beyond its initial mass of $\sim 5.1\times10^{11}\,M_\odot$ throughout its evolution, though we do not account for halo growth and triaxiality. The resulting Briggs plot for the host disk angular momentum in this case is provided in the right panel of Fig.~\ref{fig:gse_merger}. While the general properties are similar to the low-mass disk scenario---the disk precesses and eventually flips over---the details of the angular momentum evolution do change. Tilting rates follow a similar pattern, but with a lower peak tilting rate of $\sim60\,^\circ\!/\textrm{Gyr}$. Moreover, the spatial distribution of the resulting DM and stellar debris from the GSE appear different from the solar position for the two cases. This is solely due to the different disk tilting behavior (we find that the orbital evolution of the GSE-like satellite remains essentially unchanged between the two simulation runs).

Our results could have important implications for studies such as that of \cite{naidu2021reconstructing}, which seek to reconstruct merger properties of the GSE by comparing observational data to simulations. We find that small variations in the infall redshift of the GSE satellite or host disk mass can have significant effects on the stellar disk's orientation and the observed debris at $z=0$. Thus, care should be taken when comparing merger simulations to data.

Finally, one may justifiably wonder whether the large and persistent disk tilting that we observe in our GSE simulations would also be observed if the simulation were run in a cosmological context, where both the disk and halo mass grow with time and the disk's angular momentum can be affected by e.g., cosmic shear and gas accretion.  Fortunately, we have a point of comparison in the results of \cite{2022MNRAS.513.1867D}, who analyze the impact of GSE-type mergers on the stellar disks of fifteen Milky Way--like galaxies in the \textsc{Artemis} hydrodynamic cosmological simulations. They find that the angular momentum of the stellar disks in the hosts can change rapidly; the maximum averaged tilting rates for the disks in their host galaxies range from $\sim 10\,^\circ\!/\text{Gyr}$ to $\sim 60\,^\circ\!/\text{Gyr}$. \cite{2022MNRAS.513.1867D} cannot conclusively claim that the GSE-like mergers in their halos caused the disk tilting as they cannot isolate the specific effect. However, the fact that we can produce comparable maximum tilting rates with an isolated GSE-like merger, even for simplified halo densities, is a strong indication that the merger is the direct cause for the disk's evolution in the \textsc{Artemis} simulations.

\subsection{Lessons for Direct Detection}
\label{sec:direct_detection}

\begin{figure*}
    \centering
    \includegraphics[width=\textwidth]{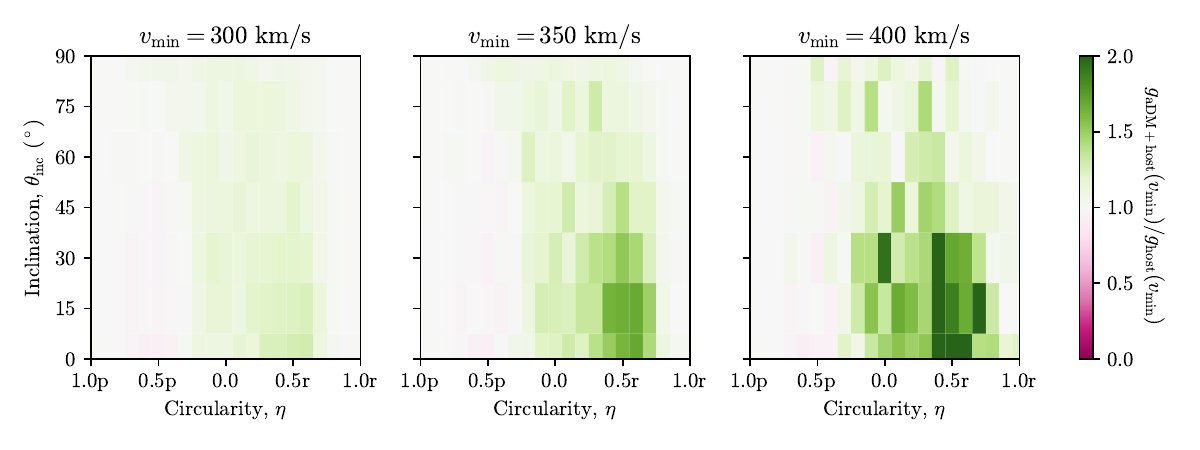}
    \caption{Ratio of $g(v_{\rm min})$ calculated with host and aDM particles to that calculated with host particles only for simulations with a $10^{11}~M_{\odot}$ satellite. The ratio is computed from particles passing through an annulus with inner radius $7.5~\text{kpc}$, outer radius $8.5~\text{kpc}$, and height $1~\text{kpc}$, centered, aligned and rotating with the host stellar disk, at the end of the simulation ($t=8~\text{Gyr}$). Each panel corresponds to different values of $v_{\rm min}$. As $v_{\rm min}$ grows, the ratio of $g(v_{\rm min})$ becomes more sensitive to enhancements of the high tail of the DM velocity distribution.}
    \label{fig:Direct_Detection}
\end{figure*}

The results of this study can be used to infer how satellite mergers in the Milky Way can potentially affect the local distribution of DM. This has consequences for direct detection experiments, which search for astrophysical DM particles scattering off of targets in terrestrial experiments. Such experiments typically utilize energy stored in the local DM distribution (at the Solar position and in the Solar rest frame) to generate a detectable signal within the detector. While the DM particle can itself never be observed in its interaction, the momentum and energy that it imparts to, say, a nucleus in the target can be detected through e.g., scintillation light or heat energy from the recoiling nucleus \citep{doi:10.1146/annurev.nucl.54.070103.181244}.

The scattering rate in a direct detection experiment is sensitive to the local DM density, $\rho_{\rm DM}$, as well as the local velocity distribution of DM, $f(v)$.  Contributions to the local phase-space density of DM from tidal debris of merging satellites can lead to particularly distinctive experimental signals \citep{2013RvMP...85.1561F}. The standard assumption for the DM density is that it is close to the value measured by e.g., performing a Jeans analysis of local stars under the assumption of equilibrium, namely $\rho_{\rm DM} \sim 0.3$ GeV/cm$^3$~\citep{2014JPhG...41f3101R}. More importantly, the standard assumption for the local velocity field is that it follows a Maxwell-Boltzmann distribution with a cut-off at the local escape velocity of the Milky Way~\citep{Baxter:2021pqo}.

Variations of these assumptions can have important implications for analyzing experimental results, depending on the specifics of the experimental setup and on the DM model being searched for. For example, the expected signal rate for any such experiment is linearly proportional to the local DM density. Additionally, the DM velocity distribution enters the signal rate calculation by modifying the kinematics of the incoming DM. For example, an enhancement from aDM particles of the high velocity tail of the host halo's velocity distribution would increase the available kinetic energy of DM particles and could significantly lower the DM mass threshold of an experiment. This could lead to a dramatic increase in the total rate for DM masses which would otherwise have an exponentially suppressed signal. For example, substantial rate enhancements from this scenario are expected for models of inelastic DM~\citep{Smith_2001,Bramante:2016rdh}. In such scenarios, a lower-mass dark particle upscatters off the nucleus to a higher-mass dark particle.

The differential scattering rate of DM with a target material, $\text{d}R/\text{d}E_r$, is proportional to the function 
\begin{equation}
g(v_{\rm min}) \equiv \int_{v_{\rm min}}^\infty \textrm{d}^3 v\, \frac{f(\mathbf{v})}{v}\,,
\end{equation}
where $E_r$ is the recoil energy of the target and $v_{\rm min}$ is the minimal velocity of incoming DM that is sufficient to produce a detectable scattering event. Its value depends on the specific experimental setup and on the kinematics. For the case of an inelastic DM scattering with a free nucleus, $v_{\rm min}$ is a function of the nucleus' mass $m_N$, the DM---nucleus reduced mass $\mu$, and the inelastic splitting of the DM model $\delta$,
\begin{equation}
v_{\rm min} = \frac{1}{\sqrt{2 E_r m_N}} \left( \frac{m_N}{\mu} E_r + \delta \right).
\end{equation}
Therefore, $v_{\rm min}$ is set by the range of recoil energies that produce a detectable signal and by the value of the inelastic splitting. As these values grow, so does $v_{\rm min}$, and therefore increased support on the high tail of $f(v)$ corresponds to an increased signal rate.

Fig.~\ref{fig:Direct_Detection} shows the ratio of $g(v_{\rm min})$ calculated with all DM particles (both host and aDM) to that calculated with host particles only, for simulations with a $10^{11}~M_{\odot}$ satellite. For a given simulation, the ratio is computed by including particles that pass through an annulus of inner radius $7.5~\text{kpc}$, outer radius $8.5~\text{kpc}$, and height $1~\text{kpc}$, centered, aligned and rotating with the host stellar disk, over the period $6$--$8$ Gyrs. This should be thought of as a proxy for the Solar position and the Solar rest frame. From left to right, the panels correspond to increasing values of $v_{\rm min}$, with each panel showing results as a function of circularity and inclination angle.

As $v_{\rm min}$ increases, the ratio of $g(v_{\rm min})$ becomes more and more sensitive to enhancements of the high tail of the velocity distribution. We find that maximal enhancements occur for retrograde mergers with $\eta \sim 0.5$ and $\theta_{\rm inc} \sim 15^\circ$. This enhancement at low inclinations is due to higher alignment between the aDM and the stellar disk at the solar position. The enhancement at retrograde circularities is due to the more boosted velocity distribution of aDM in the solar rest frame.

\section{Conclusions}
\label{sec:conclusions}

This paper investigated the impact that a merging satellite has on the tilting dynamics of a host galaxy's stellar disk. We presented a simple analytic model for the disk response, which was verified using a suite of $N$-body simulations. For the case of an isolated and isotropic Milky Way-like host, we found that non-trivial disk tilting can result from a wide range of cosmologically-motivated initial conditions for the incoming satellite.   
Such mergers torque the stellar disk, resulting in precession and alignment of the disk's angular momentum around a particular axis. Disk tilting rates are enhanced for satellites with intermediate orbital inclination angles, $\theta_{\rm inc} \sim 45^\circ$, and suppressed for satellites with orbital planes that are aligned with or orthogonal to the mid-plane of the host's stellar disk.  Additionally, tilting rates are enhanced for more massive satellites, certain values of orbital circularity, and retrograde mergers.  

We found that the host disk's overall morphology, as indicated by the co-rotation parameter, is preserved as it responds dynamically to torques from the infalling satellite and its tidal debris.  The disk itself is thickened due to the interaction with the satellite and its half-mass scale height can increase by an amount that is correlated with the details of the merger. For the example of a $\theta_{\rm inc} = 45^\circ$ orbit, we find that the scale height does not increase by more than 30\% from its original value, and this occurs only for the most radial and massive mergers considered. 

Given the disk tilting rates that we recover for cosmologically-motivated mergers, it is reasonable to expect that \emph{Gaia} may be able to detect a non-zero disk tilting rate, assuming their estimated sensitivity for this quantity~\citep{Perryman:2014gka}.  Such a measurement (or even a strong constraint) could provide an additional window into the historical evolution of the Milky Way.
This promising result motivates refining the predictions made in this work in a variety of ways. As an example, previous work such as that of~\cite{2021ApJ...919..109G} have shown that the Milky Way's DM halo responds in a non-trivial way to merger events. This suggests a followup analysis of our own simulation suite to investigate the underlying mechanisms which govern the halo's response. Additionally, important elements that can potentially be integrated into future $N$-body simulations studying disk tilting include a bulge component, a triaxial (not isotropic) DM halo, and a growing stellar disk (see e.g., \cite{2022arXiv220207662H}). By not including halo triaxiality (aligned with the disk) and stellar disk growth, it is possible that this study overestimates disk tilting rates. Moreover, it would be important to understand how the tilting rate is affected by the disk's response to multiple mergers.

Such modifications would also enable us to refine our predictions of the disk's response to the GSE merger. Our preliminary study of this question suggests that a GSE-like merger can lead to a substantial and protracted tilting response of the host disk, with it precessing and ultimately flipping over. The maximum tilting rates that we recovered from our idealized simulations are consistent with those found from GSE-like mergers in fully cosmological hydrodynamic simulations~\citep{2022MNRAS.513.1867D}.  Taken together, these results strongly suggest that the Milky Way's disk may still be evolving today due to the GSE.  Along with the simulation refinements described above, which would help make more precise predictions for the disk response to GSE, it would also be interesting to study the additional effects of the Large Magellanic Cloud.  

As a separate application of disk tilting, we explored how it correlates with non-standard properties of the DM distribution at Earth, which could impact DM detection in terrestrial experiments. In particular, we found that the high tail of the velocity distribution is most enhanced for retrograde mergers, which is particularly relevant for scattering rates of DM models with high energy scattering thresholds. Future work could make more specific predictions for mergers in the Milky Way's history, with simulation refinements as described above.

Lastly, we note that the results presented in this work are specific to CDM, and that it is a potentially interesting question to ask how the  disk tilting rates would change for alternative DM models. For example, in the case that DM has large self-interactions~\citep{2018PhR...730....1T}, DM substructure that causes baryonic disk tilting may quickly wash out. This would imply that tilting lasts for a much shorter amount of time. Models such as Fuzzy Dark Matter can affect the dynamical friction felt by the infalling satellite~\citep{Lancaster:2019mde}, and thus the distribution of its tidal debris, which can also impact the disk tilting rate. Further study will be  needed to understand whether disk tilting can provide a window into the fundamental particle nature of DM.

\section*{Acknowledgements} 
The authors are extremely grateful to L.~Necib and B.~Ostdiek for collaboration at early stages of this work.  The authors thank V.~Belokurov, N.~Garavito-Camargo, R.~Naidu, and D.~Spergel for fruitful conversations. They also acknowledge D.~Folsom for providing the \texttt{SatGen} merger trees used in this work. BD was supported in part by the Princeton Physics Department’s undergraduate
summer research program, the Princeton Class of '55 Fund, and the Princeton Class of '84 Fund. ML gratefully acknowledges financial support from the Schmidt DataX Fund at Princeton University made possible through a major gift from the Schmidt Futures Foundation. ML and OS are  supported by the DOE under Award Number DE-SC0007968 and the Binational Science Foundation (grant~No.~2018140). TC is supported by the DOE under award number DE-SC0011640. This work was performed in part at the Aspen Center for Physics, which is supported by NSF grant PHY-1607611.  ML and OS are also grateful to the organisers of the Pollica Summer Workshop supported by the Regione Campania, Universit\`{a} degli Studi di Salerno, Universit\`{a} degli Studi di
Napoli “Federico II,” i dipartimenti di Fisica “Ettore Pancini”  and “E R Caianiello,” and Istituto Nazionale di Fisica Nucleare. Vogliamo ringraziare in particolare Carmela De Bellis, Stefano Milioti, Bernardo Saviano, and Caterina Schiavo per la loro generosa ospitalit\`{a} a Pollica.  

The work presented in this paper was performed on computational resources managed and supported by Princeton Research Computing. This research made extensive use of the publicly available codes
\texttt{IPython}~\citep{PER-GRA:2007}, 
\texttt{Jupyter}~\citep{Kluyver2016JupyterN}, \texttt{matplotlib}~\citep{Hunter:2007}, 
\texttt{NumPy}~\citep{numpy:2011}, and 
\texttt{SciPy}~\citep{Jones:2001ab}.

\section*{Data Availability}

 The data underlying this article will be shared on reasonable request to the corresponding author.



\bibliographystyle{mnras}
\bibliography{main} 



\appendix

\section{A Toy Model for Disk Tilting Dynamics}
\label{sec:Disk_Dynamics}

\setcounter{equation}{0}
\setcounter{figure}{0}
\setcounter{table}{0}
\renewcommand{\theequation}{A\arabic{equation}}
\renewcommand{\thefigure}{A\arabic{figure}}
\renewcommand{\thetable}{A\arabic{table}}

This appendix presents details of a toy model that provides intuition for the possible dynamics of the stellar disk's angular momentum vector, $\vec{L}_\star$, in terms of the collective behavior of accreted dark matter~(aDM) particles orbiting in the joint potential of the host halo and the stellar disk. As discussed in Sec.~\ref{sec:disktilting} and Eq.~\eqref{eq:L_T}, the projection of $\vec{L}_\star$ onto the plane perpendicular to $\hat{z}'$ should approximately cancel the projection of  $\vec{L}_{\rm aDM}^{\rm inner}$ onto the same plane. To gain intuition, it is useful to first consider the simple case of a small number of aDM particles orbiting a \emph{fixed} stellar disk embedded within the host halo. Generalizing to the case of many aDM particles and a dynamic stellar disk is then straightforward.

In this description, there are only a small number of aDM particles, $|\vec{L}_\star| \gg |\vec{L}_{\rm aDM}|$, and therefore the $\hat{z}'$ axis is approximately perpendicular to the stellar disk. A single orbiting aDM particle, with label $i$ and mass $m_i$, orbits in a potential that has a spherical contribution and a cylindrical contribution,
\begin{equation}
    \Phi \equiv \Phi_{\rm sph}(r) + \Phi_{\rm cyl}(\rho_\star,z_\star) \,,
    \label{eq:Potential}
\end{equation}
where $\rho_\star = \sqrt{x_\star^2 + y_\star^2}$. Its change in angular momentum between any initial time $t_0$ and final time $t$ is
\begin{equation}
    \Delta \vec{L}_i
    = m_i \int_{t_0}^t a_{\star}\left(\rho_{\star,i},z_{\star,i}\right) \left(y_{\star,i}\hat{x}_\star - x_{\star,i}\hat{y}_\star\right) \text{d}\tilde{t},
    \label{eq:L_a}
\end{equation}
where the coordinates $\vec{r}_{\star,i} \equiv \{x_{\star,i}(\tilde{t}),y_{\star,i}(\tilde{t}),z_{\star,i}(\tilde{t})\}$ correspond to the $i^{\rm th}$ particle's position as a function of time (represented by the integration variable $\tilde{t}$). They can be calculated using $\ddot{\vec{r}}_{\star,i} = -\vec{\nabla} \Phi$ together with the particle's initial conditions. The function $a_{\star}(\rho_\star,z_\star)$ has units of acceleration and is given by
\begin{equation}
    a_{\star}(\rho_\star,z_\star) = - \left(\frac{\partial}{\partial z_\star} - \frac{z_\star}{\rho_\star} \frac{\partial}{\partial \rho_\star} \right) \Phi_{\rm cyl}(\rho_\star,z_\star) \, .
\end{equation}

From Eq.~\eqref{eq:L_a}, it is clear that $\Delta\vec{L}_i \cdot \hat{z}_\star$ is zero, however $\Delta\vec{L}_i \cdot \hat{x}_\star$ and $\Delta\vec{L}_i \cdot \hat{y}_\star$ are in general non-zero. They depend on the specific form of $\Phi_{\rm cyl}$ and on the orbit of the particle. Eq.~\eqref{eq:L_a} can easily be solved numerically, however a simplification occurs since the numerical solution is well approximated by the following function of $\Delta t \equiv t - t_0$,
\begin{equation}
    \Delta \vec{L}_i = L_{i}^{T} \left[\vec{\mathcal{P}}(\omega_i,\varphi_i,\Delta t) - \vec{\mathcal{P}}(\omega_i,\varphi_i,0)\right] 
    \label{eq:L_i_approx}
\end{equation}
with the simple periodic function,
\begin{equation}
    \vec{\mathcal{P}}(\omega,\varphi,\Delta t) \simeq \text{Re}\left[e^{i \omega \Delta t} e^{i\varphi}\right] \hat{x}_\star + \text{Im} \left[e^{i \omega \Delta t} e^{i\varphi}\right] \hat{y}_\star \,.
    \label{eq:P_def}
\end{equation}
The periodic function has been written in complex form instead of using sin and cos notation for later convenience. This corresponds to an approximate precession of the particle's angular momentum around the $\hat{z}_\star$ axis. In the above equations, $L_i^T$ is the projection of the particle's angular momentum onto the $x_\star-y_\star$ plane and $\varphi_i$ is a phase set by the initial conditions of the particle. The value of the frequency $\omega_i$ depends on both the initial conditions of the particle and on the precise form of the potential $\Phi$.

Using the above equations, one can calculate the sum of angular momenta change for $N$ orbiting aDM particles, which can be thought of as those contributing to $\vec{L}_{\rm aDM}^{\rm inner}$,
\begin{equation}
    (\Delta \vec{L}_{\rm aDM}^{\rm inner})^T = \sum_i \Delta \vec{L}_i \equiv N \big\langle \Delta \vec{L} \big\rangle \,,
    \label{eq:L_disk}
\end{equation}
where $\langle...\rangle$ denotes an average over all particles. This average is straightforward to calculate if one knows the distributions of $L_i^T$, $\omega_i$ and $\varphi_i$, and noting that an additional simplification occurs if the amplitude $L_i^T$ is highly correlated with $\omega_i$. Then, the discrete values of angular momentum for each particle become a continuous function $\Delta \vec{L}_i \to \Delta \vec{L} (\omega, \varphi, \Delta t) = L^{T}(\omega) [\vec{\mathcal{P}}(\omega,\varphi,\Delta t) - \vec{\mathcal{P}}(\omega,\varphi,0)]$. If the probability distribution function of frequencies and phases $f(\omega,\varphi)$ is known, the result can be rewritten as
\begin{equation}
   \big\langle \Delta \vec{L} \big\rangle \simeq \int \Delta \vec{L}(\omega, \varphi, \Delta t) f(\omega,\varphi) \text{d}\omega \text{d}\varphi \,,
   \label{eq:Sum_L_Fourier}
\end{equation}
where the probability function is normalized to unity, $\int f(\omega,\varphi)\text{d}\omega \text{d}\varphi = 1$. (This is essentially just the result of a Fourier analysis.) 

\begin{figure*}
    \centering
    \includegraphics{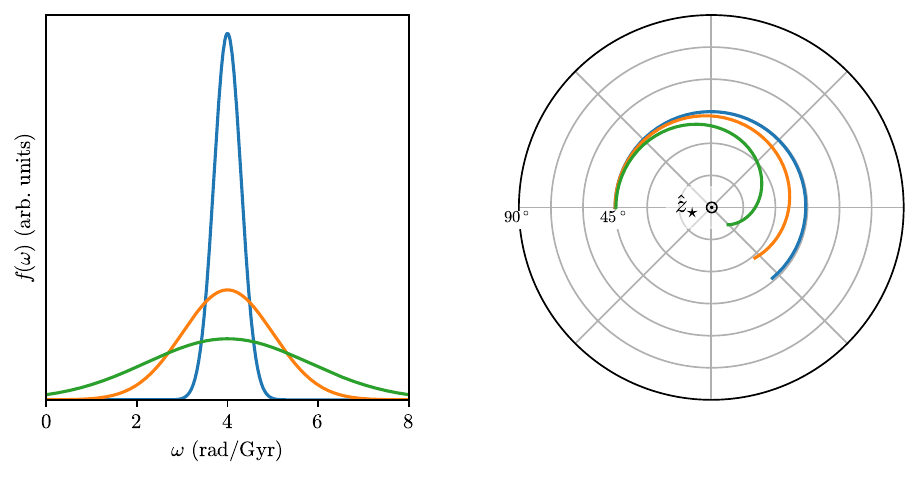}
    \caption{Briggs plot showing the expected angular momentum evolution of aDM particles (and therefore also of the stellar disk since $(\vec{L}_\star)^T \simeq - (\vec{L}_{\rm aDM}^{\rm inner})^T$) based on Eqs.~\eqref{eq:L_disk} and~\eqref{eq:Sum_L_Fourier}. Each curve corresponds to a different distribution $f(\omega)$ under the assumption that all particles have $\varphi=0$ (shown in the left panel). The blue curve shows an example of a tight Gaussian distribution which produces mostly precession motion around $\hat{z}_\star$. The orange and green curves are also Gaussian, but with increasing widths. These produce precession motion coupled with alignement of the aDM angular momentum vector with the $\hat{z}_\star$ axis.}
    \label{fig:Simple_Model}
\end{figure*}

As a simple example, consider the case where $f(\omega,\varphi)$ is the product of two Gaussians,
\begin{equation}
    f(\omega,\varphi) = \frac{1}{2\pi \sigma_\omega \sigma_\varphi} e^{-\frac{1}{2}\left[\left(\frac{\omega - \bar{\omega}}{\sigma_\omega}\right)^2 + \left(\frac{\varphi-\bar{\varphi}}{\sigma_\varphi}\right)^2\right]},
\end{equation}
and $L^{T}(\omega)$ does not vary appreciably in the range where the function has support. Then Eq.~\eqref{eq:Sum_L_Fourier} can be evaluated to find
\begin{equation}
    (\Delta \vec{L}_{\rm aDM}^{\rm inner})^T \simeq N L^{T}(\bar{\omega}) e^{-\frac{1}{2}\sigma_\varphi^2} e^{i \bar{\varphi}} \left[ e^{-\frac{1}{2}  \sigma_\omega^2 \Delta t^2} e^{i\bar{\omega} \Delta t} - 1 \right],
    \label{eq:L_Gaussian}
\end{equation}
where it should be understood that the real part of the right-hand side is in the $\hat{x}_\star$ direction and the imaginary part in the $\hat{y}_\star$ direction, as in Eq.~\eqref{eq:P_def}.

The result in Eq.~\eqref{eq:L_Gaussian} provides intuition for the dynamics of $(\vec{L}_\star)^T \simeq - (\vec{L}_{\rm aDM}^{\rm inner})^T$. First, remember that $\hat{z}_\star$ is close to, but not precisely equal to $\hat{z}'$. As $|\vec{L}_{\rm aDM}|$ grows, these directions diverge and the true motion of both aDM particles and stellar disk angular momenta will be around the direction $\hat{z}'$. If the spreads in frequencies and phases ($\sigma_\omega$ and $\sigma_\varphi$,  respectively) are both small, then only the complex arguments of the exponents survive and the stellar disk reacts by precessing around the $\hat{z}'$ direction at a frequency $\bar{\omega}$. If $\sigma_\varphi$ is sizable, the amplitude of precession is exponentially suppressed by the square of its value. If $\sigma_\omega$ is non-zero, the amplitude of precession becomes exponentially more suppressed over time at a rate set by the square of its value. This time evolution corresponds to an alignment of the stellar disk towards the axis of precession. Importantly, these two types of dynamics---precession around the $\hat{z}'$ axis and alignment with that axis---are precisely those observed in our simulations.

Of course, in a realistic scenario the approximations made above need not hold. Specifically, the assumptions that $|\vec{L}_\star| \gg |\vec{L}_{\rm aDM}|$, that the potential is time-independent, and that the distributions of $\omega_i$ and $\varphi_i$ are Gaussian need not be correct. As already discussed, relaxing the assumption that $|\vec{L}_\star| \gg |\vec{L}_{\rm aDM}|$ just corresponds to changing the direction of $\hat{z}'$ away from the perpendicular to the stellar disk, but otherwise does not change the qualitative picture. Regarding the time dependence of the potential, the resulting behavior is still straightforward to understand since there is always some finite timescale with which the stellar disk reacts. For shorter times, Eq.~\eqref{eq:L_Gaussian} approximately holds. Over longer times, one can simply think of the dynamics as short time-step evolutions of $(\vec{L}_\star)^T$ according to Eq.~\eqref{eq:L_Gaussian} with some values of $\{\bar{\omega}, \sigma_{\omega}, \bar{\varphi}, \sigma_{\varphi}\}$, and at every new time-step these values simply evolve as well. Finally, the assumption of a non-Gaussian form for $f(\omega,\varphi)$ renders Eq.~\eqref{eq:L_Gaussian} non-analytical, but does not change the qualitative picture.

Figure~\ref{fig:Simple_Model} shows three examples of the evolution of $\vec{L}_{\rm aDM}^{\rm inner}$ based on Eqs.~\eqref{eq:L_disk} and~\eqref{eq:Sum_L_Fourier}. Each color corresponds to a different distribution for $f(\omega)$, taking all particles to have $\varphi=0$, also shown as a separate panel. All three distributions are Gaussian and each curve corresponds to a different width, $\sigma_\omega$. The behavior described above is evident---namely the aDM's angular momentum (and therefore also the stellar disk's) precesses around the $\hat{z}_\star$ axis (which is equal to $\hat{z}'$ in this case) and also aligns with it at a rate that grows as $\sigma_\omega$ grows. This figure should be compared to Fig.~\ref{fig:Prec_Align_Toy}, where the average angular momentum was calculated numerically over time for an ensemble of aDM particles (with either a large or small spread in $\omega$) orbiting within the potential of a stellar disk and an NFW halo. As can be seen, the toy model presented in this section qualitatively reproduces the behavior observed in Fig.~\ref{fig:Prec_Align_Toy}.

\section{Supplementary Figures}
\label{sec:supp_figures}

In this appendix, we provide summary results for our simulations. Fig.~\ref{fig:app_highres_polar} shows an additional high-resolution run, exhibiting interesting behavior before satellite disruption. Fig.~\ref{fig:app_tilting} shows the tilting rate for all combinations of mass, circularity, and inclination, measured over two 1 Gyr time intervals. Fig.~\ref{fig:app_thickening} shows the degree of disk thickening as measured by $z_{1/2}$, again for the entire parameter space. Tab.~\ref{tab:app_highres} presents tilting rates for all of our high-resolution simulations, showing agreement with results from low-resolution runs. These figures demonstrate the robustness of the trends described in Sec.~\ref{sec:Results}. 

\setcounter{equation}{0}
\setcounter{figure}{0}
\setcounter{table}{0}
\renewcommand{\theequation}{B\arabic{equation}}
\renewcommand{\thefigure}{B\arabic{figure}}
\renewcommand{\thetable}{B\arabic{table}}

\begin{figure}
    \centering
    \includegraphics{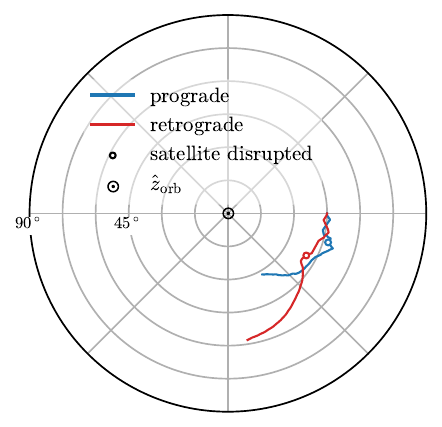}
    \caption{Evolution of the host stellar disk's angular momentum for high-resolution simulations of a $10^{11} M_\odot$ satellite with $\eta = 0.3$ and $\theta_{\rm inc}=45^\circ$ on a prograde~(blue) or retrograde~(red) orbit. This complements Fig.~\ref{fig:Pro_Retro_Toy} with lower circularity and Fig.~\ref{fig:polar_plot} with higher resolution. Prior to disruption, the host disk wobbles due to torques from the infalling bound satellite. Note the opposite direction of these oscillations for prograde and retrograde mergers.}
    \label{fig:app_highres_polar}
\end{figure}

\begin{figure*}
    \centering
    \includegraphics[width=\textwidth]{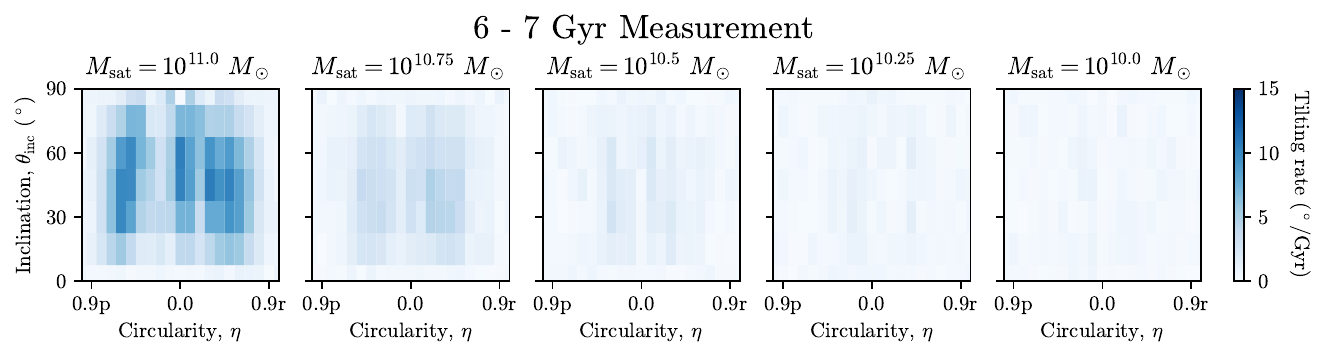}
    \includegraphics[width=\textwidth]{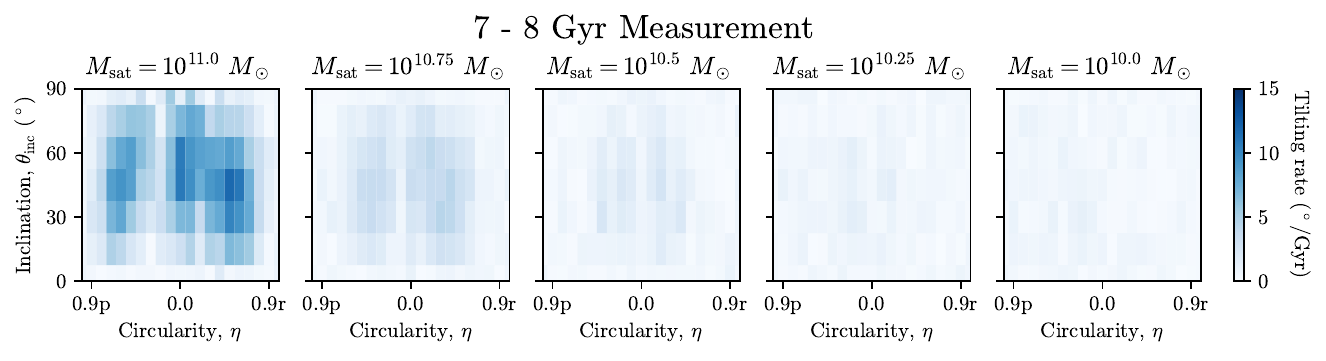}
    \caption{Disk tilting rates for all combinations of satellite mass, inclination, and circularity at low resolution. Measurements taken from two 1 Gyr intervals are shown to demonstrate robustness. The key trends are also presented in Fig. \ref{fig:tip_precession}. For satellites below $10^{10.5} M_\odot$, disk tilting rates are below the noise. Note that the magnitude of disk tilting should not be taken as a precise prediction due to the idealized nature of our simulations.}
    \label{fig:app_tilting}
\end{figure*}
    
\begin{figure*}
    \centering
    \includegraphics[width=\textwidth]{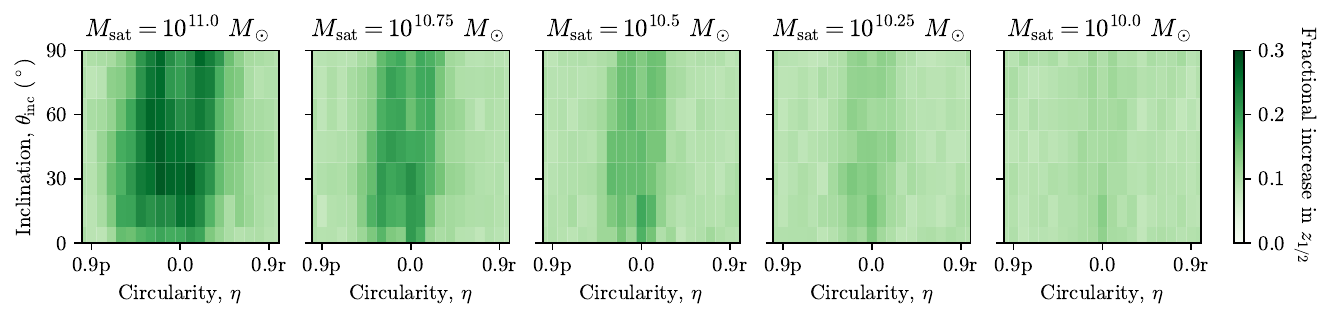}
    \caption{Fractional increase in half-mass height of the host disk from the initial time to the final time of the simulation ($t=8~\text{Gyr}$). Results are presented for all low-resolution simulations. The dependence on inclination is less dramatic than the dependence on circularity and satellite mass, as presented in Fig. \ref{fig:disk_heating}.}
    \label{fig:app_thickening}
\end{figure*}

\begin{table*}

    \centering
    \begin{tabular}{lllllll}

\hline
\hline
&&& \multicolumn{4}{c}{High (low) resolution tilting rate ($^\circ/\text{Gyr}$)} \\
Inclination, $\theta_{\rm inc}$ ($^\circ$) & Circularity, $\eta$ & Direction & \multicolumn{2}{l}{$6-7~\text{Gyr}$}\hspace{40pt} & \multicolumn{2}{l}{$7-8~\text{Gyr}$} \\
\hline

45 & 0.5 & prograde   & 9.2 & (9.8)  & 8.1 & (8.4) \\
45 & 0.5 & retrograde & 9.9 & (10.2) & 10.8 & (11.7) \\
45 & 0.3 & prograde   & 5.2 & (4.5)  & 4.8 & (4.3) \\
45 & 0.3 & retrograde & 10.1 &(10.6) & 9.1 & (9.0) \\
45 & 0.0 &   n/a      & 10.2& (10.2) & 10.4 & (10.8) \\
90 & 0.5 &   n/a      & 3.0 & (3.1)  & 1.1 & (1.1) \\
90 & 0.3 &   n/a      & 1.3 & (0.9)  & 0.9 & (0.3) \\

\hline
\hline

    \end{tabular}
    
\caption{ Merger parameters for 7 high resolution ($m_p = 10^{4.5}~M_\odot$) simulations. These were chosen to cover important regions of the parameter space. All simulations have $M_{\rm sat}=10^{11}~M_\odot$. Tilting rates for two 1 Gyr intervals are provided for comparison with low resolution ($m_p=10^{5.5}~M_\odot$) runs. We find that results are qualitatively consistent between the two. }\label{tab:app_highres}
\end{table*}


\bsp	
\label{lastpage}
\end{document}